\documentclass[letterpaper,titlepage,11pt]{article}

\usepackage[T1]{fontenc}
\usepackage[utf8]{inputenc}
\usepackage{lmodern}
\usepackage{enumerate}
\usepackage[english]{babel}
\usepackage{comment}
\usepackage[numbers,sort&compress]{natbib}
\usepackage{bm}
\usepackage[usenames,dvipsnames]{xcolor}
\usepackage[a4paper]{geometry}
\usepackage[normalem]{ulem}

\usepackage[colorlinks=true,urlcolor=blue,citecolor=magenta]{hyperref}
\usepackage{amsmath,bbm}
\usepackage{amsfonts}
\usepackage{amssymb}
\usepackage{graphicx}
\usepackage[dvipsnames]{xcolor}
\usepackage{mathtools}
\usepackage{simplewick}
\usepackage{hyperref}
\usepackage{multirow}
\usepackage{enumitem}
\usepackage{mathrsfs}
\usepackage{physics}
\usepackage[compat=1.1.0]{tikz-feynman}
\usepackage{subcaption}
\usepackage{booktabs}
\usepackage{xcolor}

\usepackage{lipsum}

\usepackage{amsthm}

\allowdisplaybreaks[1]

\newcommand\blfootnote[1]{%
  \begingroup
  \renewcommand\thefootnote{}\footnote{#1}%
  \addtocounter{footnote}{-1}%
  \endgroup
}

\newcommand{\Z}{\mathbb{Z}}
\newcommand{\C}{\mathbb{C}}

\newcommand{\CN}{\mathcal{N}}

\newcommand{\CO}{\mathcal{O}}

\newcommand{\nn}{\nonumber}
\newcommand{\spa}{\ , \ \ }

\hypersetup{   colorlinks=true,  citecolor=blue, linkcolor=blue, urlcolor=red }

\begin{document}

\numberwithin{equation}{section}

\begin{titlepage}
\rightline{\vbox{   \phantom{ghost} }}

 \vskip 1.8 cm
\begin{center}
{\LARGE \bf
Hagedorn temperature 
from the thermal scalar \\[2mm] in AdS and pp-wave backgrounds}
\end{center}
\vskip 1 cm

\title{}
\date{\today}
\author{Troels Harmark}

\centerline{\large {{\bf Troels Harmark$^1$}}}

\vskip 1.0cm

\begin{center}
\sl ${}^2$Niels Bohr International Academy, Niels Bohr Institute, University of Copenhagen,\\
Blegdamsvej 17, DK-2100 Copenhagen \O, Denmark\\[1mm]
\end{center}

\vskip 1.3cm \centerline{\bf Abstract} \vskip 0.2cm 
\noindent
We propose a thermal scalar equation of motion (EOM) that takes into account curvature corrections for backgrounds supported by Ramond-Ramond fluxes. 
This can be used to obtain the Hagedorn temperature for type II string theory on AdS and pp-wave backgrounds.
For Ramond-Ramond flux supported pp-waves we show that the proposed thermal scalar EOM reproduces the leading curvature correction in the Hagedorn temperature equation obtained from the type II string theory spectrum.
Furthermore, we use the thermal scalar EOM to obtain curvature corrections to the Hagedorn temperature for the AdS$_5 \times S^5$ and AdS$_4 \times \C P^3$ backgrounds. These corrections match with strong coupling results of the integrable dual field theories, recently obtained by the Quantum Spectral Curve technique.

\blfootnote{ \scriptsize{ \texttt{harmark@nbi.ku.dk}} }

\end{titlepage}
\newpage
\tableofcontents

%%%%%%%%%%%%%%%%%%%%%%%%%%%%%%%%%%%%%%%%%%%%%%%%%%%
\section{Introduction}

The Hagedorn temperature is a source of continuing fascination and insight into thermal physics of strings and gauge theories. 
Originally proposed in QCD \cite{Hagedorn:1965st}, the exponential growth of states associated to the Hagedorn temperature was found to be a feature of string theory as well. The deeper reason for this 
shared feature is now seen as a consequence of the duality between gauge and string theories, as first made precise by the AdS/CFT correspondence \cite{Maldacena:1997re}. In its most studied form, this is a duality between SU$(N)$ $\mathcal{N}=4$ superconformal Yang-Mills (SYM) theory on $\mathbb{R} \times S^3$ and type IIB string theory on AdS$_5 \times S^5$.
Through this, the Hagedorn temperature is connected to the confinement/deconfinement phase transition on the gauge theory side, and the Hawking-Page phase transition on the string theory side \cite{Sundborg:1999ue,Aharony:2003sx}. 
The Hagedorn behavior of the spectrum arises at high energies in the limit of a large number of colors on the gauge theory side corresponding to having weak string coupling on the string theory side.

As discovered in \cite{Sundborg:1999ue,Aharony:2003sx}, one can compute the Hagedorn temperature on the gauge theory side to zeroth order in the 't Hooft coupling $\lambda$
from analyzing the partition function. This analysis was later extended to the one-loop correction in $\lambda$ \cite{Spradlin:2004pp}. 
However, to connect to the string side of the duality one needs to find the strong-coupling limit of the Hagedorn temperature. 
A first step towards this goal was taken in \cite{Harmark:2017yrv} where it was found how to compute the Hagedorn temperature using the integrability symmetry of $\mathcal{N}=4$ SYM theory \cite{Beisert:2010jr,Bombardelli:2016rwb}, in part employing an earlier technique of \cite{Harmark:2006ta} for a decoupled sector of the theory.
Building on this,  \cite{Harmark:2018red,Harmark:2021qma} showed how to compute the Hagedorn temperature analytically to seven loop order, as well as numerically from zero to $\lambda=1660$, using the Quantum Spectral Curve (QSC) technique \cite{Gromov:2013pga,Gromov:2014bva,Gromov:2014caa}. This made it possible to reach the asymptotic Hagedorn temperature at strong coupling where it should match the flat space Hagedorn temperature of type IIB string theory, $T_{\rm H} = \tfrac{1}{\sqrt{8} \pi l_s}$ with $l_s$ being the string length. Measured in units of the $S^3$ radius, and introducing $g^2 = \tfrac{\lambda}{16\pi^2}$, this can be written as $T_{\rm H} = \sqrt{\tfrac{g}{2\pi}}$. This is the limiting value of the Hagedorn temperature for $g\rightarrow \infty$.
The numerical QSC results for $\CN=4$ SYM were subsequently improved in \cite{Ekhammar:2023glu} and the Hagedorn temperature calculation was extended to the ABJM theory as well in \cite{Ekhammar:2023cuj}.

Recently, important advances have been made on the string theory side, with the discovery that the so-called {\sl thermal scalar} can compute not just the leading order flat-space temperature at infinite coupling, but also the first order correction \cite{MaldacenaPrivateCommunication,Urbach:2022xzw}, giving $T_{\rm H} \simeq \sqrt{\tfrac{g}{2\pi}} + \tfrac{1}{2\pi}$, for large $g$, which matches the results of \cite{Harmark:2021qma,Ekhammar:2023glu}. 
Indeed, we can write the thermal scalar equation of motion (EOM) as
\begin{equation}
\label{intro_EOM}
\nabla_{\rm e}^{\, 2} \chi = m^2 \chi 
\spa m^2 = -\frac{2}{l_s^2} + \frac{\beta^2 G_{00}}{4\pi^2 l_s^4} \,,
\end{equation}
with $\beta=1/T_{\rm H}$, $\nabla_{\rm e}^{\, 2}$ being the Laplacian, and $G_{00}$ the metric component along the thermal circle.
The thermal scalar corresponds to the winding mode along the Euclidean time direction, which becomes tachyonic as one reaches the Hagedorn temperature \cite{Sathiapalan:1986db,Kogan:1987jd,Atick:1988si}. One can argue for this by employing a modular transformation of the torus string amplitude \cite{Barbon:2004dd,Kruczenski:2005pj}.
For type IIB string theory on AdS$_5 \times S^5$, the thermal scalar EOM \eqref{intro_EOM} captures the above-mentioned leading term and the first-order correction to $T_{\rm H}$ \cite{MaldacenaPrivateCommunication,Urbach:2022xzw}. Relative to the leading piece, the first-order correction is proportional to $l_s/\mathcal{R} \sim 1/\sqrt{g}$ where $\mathcal{R}$ is the curvature length scale of the background.

The main goal of this paper is to find the modified thermal scalar EOM which can capture higher-order corrections in $T_{\rm H}$ at orders $l_s^2/\mathcal{R}^2$ and $l_s^3/\mathcal{R}^3$ for Ramond-Ramond flux supported backgrounds of type II string theory. 
This is motivated by the recent interesting works \cite{Ekhammar:2023glu,Bigazzi:2023hxt,Ekhammar:2023cuj} that were the first to consider such corrections for AdS backgrounds. 
Backgrounds with Ramond-Ramond fluxes are especially interesting, since it seems that once a Ramond-Ramond flux is turned on, the thermal scalar EOM \eqref{intro_EOM} is no longer sufficient to capture the near-critical behavior close to the Hagedorn temperature.

As a stepping stone towards our main goal, we point out that one can use the thermal scalar to compute the Hagedorn temperature in the supergravity limit of string theory, not just for flat space and AdS backgrounds, but also for pp-wave backgrounds. Ramond-Ramond supported pp-wave backgrounds are particularly useful to consider, due to the fact that one can find exact expressions of the Hagedorn temperature as computed from the string spectrum partition function. Thus, having the possibility to compare to the exact Hagedorn temperature greatly improve the testing ground for the modified thermal scalar EOM that we seek in this paper.

The corrections in $T_{\rm H}$ at orders $l_s^2/\mathcal{R}^2$ and $l_s^3/\mathcal{R}^3$ are the first ones for which the thermal scalar EOM are sensitive to the curvature of the background. 
We approach the problem of how to modify the thermal scalar EOM by demanding that one should be able to write any curvature correction in a covariant form so that it holds for any background. 
Thus, since we have extended our family of Ramond-Ramond flux supported backgrounds to include both pp-waves and AdS spaces of various dimensions, the same curvature corrections should work for all such backgrounds. We succeed in finding two such curvature correction terms for the thermal scalar that both involve the Ricci curvature tensor for the background, and which can reproduce the known results for both AdS spaces and pp-waves. 

One of the curvature correction terms is part of the mass of the thermal scalar. It is proportional to $\beta^2 R_{00}$ in the notation of \eqref{intro_EOM} where $R_{00}$ is the component of the Ricci tensor along the time direction. We write a fully covariant version below. Considering a particular family of pp-wave backgrounds, the $\beta^2 R_{00}$  term matches the leading correction in the curvature-expanded Hagedorn temperature as computed from the type II string partition function. This fixes the coefficient of the  $\beta^2 R_{00}$  term.
For type II string theory on AdS spaces, one obtains a unique prediction for the  $l_s^2/\mathcal{R}^2$ correction to the Hagedorn temperature. We compare successfully this correction to the numerical predictions of the QSC technique for type IIB string theory on AdS$_5\times S^5$ \cite{Harmark:2018red,Harmark:2021qma,Ekhammar:2023glu}  and for type IIA string theory on AdS$_4\times \C \mbox{P}^3$ \cite{Ekhammar:2023cuj}. It also matches the string zero point energy calculation of \cite{Bigazzi:2023hxt} for AdS spaces.

The second curvature correction term instead modifies the left-hand side of \eqref{intro_EOM}. This term turns out to give zero in case of the Ramond-Ramond supported pp-wave backgrounds. Instead, it is non-zero on the AdS spaces and contributes at order $l_s^3/\mathcal{R}^3$.
Based on this, we propose the exact coefficient of the second curvature correction term by comparing to the numerical QSC predictions for the $l_s^3/\mathcal{R}^3$ correction to $T_{\rm H}$ as computed in \cite{Harmark:2018red,Harmark:2021qma,Ekhammar:2023glu,Ekhammar:2023cuj}. 

As mentioned, this paper is motivated by the recent works \cite{Ekhammar:2023glu,Bigazzi:2023hxt,Ekhammar:2023cuj} that were the first to consider 
higher-order corrections in $T_{\rm H}$ at orders $l_s^2/\mathcal{R}^2$ and $l_s^3/\mathcal{R}^3$ specifically for the AdS$_5\times S^5$ and AdS$_4\times \C \mbox{P}^3$ backgrounds.
To match the numerical QSC predictions of \cite{Harmark:2018red,Harmark:2021qma,Ekhammar:2023glu,Ekhammar:2023cuj},
the authors of \cite{Ekhammar:2023glu,Ekhammar:2023cuj} initiated an ad hoc method for these specific AdS backgrounds by modifying the thermal scalar mass square $m^2$ with a correction term
\begin{equation}
\label{adhoc_term}
\beta^2 \Delta C 
\end{equation}
with $\Delta C$ a background-dependent constant. 
Subsequently $\Delta C$ was computed exactly for all AdS$_{d+1}$ backgrounds in \cite{Bigazzi:2023hxt} by calculating the string zero point energy up to order $l_s^2/\mathcal{R}^2$, and found to be proportional to the dimension $d$. 

However, an important issue with term \eqref{adhoc_term} correcting $m^2$ is that it lacks covariance.
Specifically, it does not take into account the proper length of the thermal circle. 
This is particularly evident for AdS backgrounds, for which the proper length is coordinate dependent since $g_{00}$ depends on the AdS radius.%
\footnote{See the end of Section \ref{sec:ads} for a careful discussion.}
One can also see the lack of covariance by considering rescalings of the time coordinate. 
Moreover, the lack of covariance is the reason why $\Delta C$ is background dependent.  
Instead, both the term $\beta^2 R_{00}$, as well as the other curvature correction term in the thermal scalar EOM mentioned above, solve this issue since both terms are fully covariant. 
In addition to including the correct dependence on the proper length of the thermal circle, it also means one needs only to compute the coefficient of $\beta^2 R_{00}$  term for a single background to be able to predict the order $l_s^2/\mathcal{R}^2$ correction for all backgrounds, including not only AdS spaces but also pp-waves.

The paper is structured as follows. In Section \ref{sec:thermal_scalar} we 
introduce the thermal scalar as a method to compute the Hagedorn temperature for type II string theory backgrounds. We exhibit how to use it to compute the Hagedorn temperature for Lorentzian signature backgrounds and we review how the formalism is modified in case the background is non-static and/or with Kalb-Ramond flux. These advances are particularly useful when applying the thermal scalar EOM to pp-wave backgrounds. 
All this leads up to our conjectured new form for the curvature-corrected thermal scalar EOM in the case of type II string theory backgrounds with Ramond-Ramond flux with two new curvature-correction terms. As part of this, we discuss in general what possible curvature corrections one can have. 

In Section \ref{sec:ppwave} we consider the Hagedorn temperature for pp-wave backgrounds of type II string theory. We focus mostly on Ramond-Ramond flux supported pp-waves. We review the Hagedorn temperature as computed from the string spectrum for such backgrounds, and its expansion in powers of the curvature of the background. 
Subsequently, we exhibit the Hagedorn temperature as found from the thermal scalar formalism, and see how the curvature correction of the thermal scalar matches what one obtains from the string spectrum. This fixes one of the coefficients of the two curvature-correction terms of the thermal scalar EOM. We also discuss the extension to having a chemical potential and the case of a pp-wave with both Ramond-Ramond and Kalb-Ramond fluxes.

In Section \ref{sec:ads} we consider Ramond-Ramond flux supported backgrounds of type II string theory with an AdS part which are protected against $\alpha'$ corrections and describe the holographic dual of superconformal field theories. We focus on the cases of  AdS$_5 \times S^5$ and AdS$_4 \times \C P^3$.
We review the recent progress for computing the Hagedorn temperature in the two cases by using the QSC equation for the dual integrable superconformal field theories. This provides numerical predictions of the asymptotic Hagedorn temperature at strong coupling, which is the regime in which the thermal scalar is valid. Next we turn to the predicted Hagedorn temperature from the thermal scalar, as expanded in powers of the curvature. We show that by fixing the value of the second coefficient of the  two curvature-correction terms of the thermal scalar EOM one can obtain the correct expanded Hagedorn temperature. In particular, this shows that the same curvature-corrected thermal scalar EOM can provide the correct Hagedorn temperature for both pp-wave and AdS backgrounds. 
Finally, we explain in detail the statement that the ansatz for the correction of the thermal scalar EOM of \cite{Ekhammar:2023glu,Bigazzi:2023hxt,Ekhammar:2023cuj} does not take into account the proper length of the thermal circle.

In Section \ref{sec:discussion} we conclude the paper with a discussion of our results and possible future directions, including adding higher-order terms in the curvature to the thermal scalar EOM.

%%%%%%%%%%%%%%%%%%%%%%%%%%%%%%%%%%%%%%%%%
\section{Thermal scalar  for a general type II background}
\label{sec:thermal_scalar}

It is well-established that one can model the near-critical behavior of string theory close to the Hagedorn temperature by a single complex scalar field, known as the {\sl thermal scalar} \cite{Sathiapalan:1986db,Kogan:1987jd,Atick:1988si}. 
Close to the Hagedorn temperature, the dominant configuration of the partition function for a gas of strings in flat space with vanishing string coupling turns out to be a single long string, assuming at least one infinitely extended spatial direction \cite{Mitchell:1987hr,Mitchell:1987th,Bowick:1989us}.
One can equivalently see the Hagedorn singularity from the string path integral with a compactified euclidean time direction \cite{Polchinski:1985zf} and with strings having winding number $\pm 1$. 
By a modular transformation, the near-critical ultraviolet behavior can be mapped to a near-critical infrared behavior of a single string mode with winding mode $\pm 1$, being the thermal scalar $\chi$ \cite{Barbon:2004dd,Kruczenski:2005pj}. $\chi$ is a complex scalar corresponding to winding $1$, while $\chi^*$ corresponds to winding $-1$. The Hagedorn temperature is mapped by the modular transformation to the point when $\chi$ becomes massless. Above the Hagedorn temperature $\chi$ is tachyonic. This is in particular true for type II string theory, in which the occurrence of a tachyonic mode is due to the supersymmetry breaking boundary conditions of a thermal circle. 
The mapping of near-Hagedorn behavior to the dynamics of the thermal scalar has been argued to extend to curved space \cite{Horowitz:1997jc,Barbon:2004dd,Kruczenski:2005pj,Mertens:2013pza}. 
Moreover, one has used the thermal scalar to examine the transition between a gas of strings and black holes for non-zero string coupling, being able to model the backreaction of strings  \cite{Atick:1988si,Horowitz:1997jc}.

For the purposes of this paper, we utilize the thermal scalar to predict the occurrence of the Hagedorn temperature of type II string theory in curved space, at least for the first few orders in an expansion of small curvature. Thus, we will neglect the string coupling and backreaction effects throughout the paper.
Note also that we shall consider only type II supergravity backgrounds that are exact with respect to $\alpha'$ corrections.

%%%%%%%%%%%%%%%%%%%%%%%%%%%%%%%%%%%%%%%%%
\subsection{Thermal scalar for Euclidean section}

Consider type II string theory in ten dimensions at finite temperature $T= 1/\beta$. 
We assume a vanishing string coupling since we are interested in obtaining the Hagedorn temperature. As part of this, the dilaton field is assumed to be constant such that it decouples from the dynamics. 
The background is naturally required to be stationary.
To begin, we assume that no Kalb-Ramond or Ramond-Ramond fluxes are turned on. 

As the thermal scalar arise as a mode in the Euclidean section it is natural to formulate using the Euclidean section metric. We write
\begin{equation}
ds_{\rm e}^{\, 2} = G_{00} (dt_{\rm E})^2 + G_{ij} dx^i dx^j \,,
\end{equation}
where $i,j=1,2,...9$ and $t_{\rm E}$ is the Euclidean time. Then the thermal scalar 
 $\chi$ has the equation of motion (EOM) \cite{Horowitz:1997jc}
\begin{equation}
\label{chi_eq_E}
\nabla_{\rm e}^{\, 2} \chi = m^2 \chi \,,
\end{equation}
where the Laplace operator $\nabla_{\rm e}^{\, 2}$ is computed from the nine-dimensional metric $G_{ij}$. 
For type II string theory we have the leading contribution 
\begin{equation}
\label{m_square_orig}
m^2 =- \frac{2}{l_s^2}+ \frac{\beta^2 G_{00}}{4\pi^2 l_s^4}  \,.
\end{equation}
The Euclidean time coordinate is periodic with period $\beta$ being the inverse temperature. 
The $-2/l_s^2$ term arise from integrating out the massive string modes. It reflects the fact that the thermal circle breaks the supersymmetry of type II string theory, providing a tachyonic mode when $\beta$ is sufficiently small.
We see that $m^2=0$ for flat space correctly reproduces the Hagedorn temperature $\beta = 2\sqrt{2} \pi l_s$.

It is important to emphasize that $\beta$ naturally appears in the combination $\beta^2 G_{00}$ in $m^2$. Indeed, $\beta \sqrt{G_{00}}$ corresponds to the proper circumference of the thermal circle. Another way to argue this is by considering a rescaling of the Euclidean time
\begin{equation}
\label{time_rescaled_E}
t'_{\rm E} = \kappa \, t_{\rm E}  \spa \beta' = \kappa \, \beta \,,
\end{equation}
with $\kappa$ an arbitrary real number,
then the inverse temperature should rescale accordingly 
since $\beta$ is the periodicity of the Euclidean time. With this rescaling one finds $G'_{00} = \kappa^{-2} G_{00}$ hence $\beta'{}^2 G'_{00} = \beta^2 G_{00}$ is invariant, which is why one needs this combination when considering a more general background target space-time.

One can generalize the above equations of motion \eqref{chi_eq_E} to non-static backgrounds \cite{Mertens:2013pza}
\begin{equation}
\label{metric_eucl}
ds_{\rm E}^2 = G_{00} (dt_{\rm E})^2 + 2 G_{0i} dt_{\rm E} dx^i + G_{ij} dx^i dx^j \,.
\end{equation}
From a nine-dimensional point of view the Laplace operator $\nabla_{\rm e}^{\, 2}$ then follows by using the determinant and the inverse matrix of the 9 by 9 matrix $G_{ij} - G_{0i}G_{0j}/G_{00}$ \cite{Mertens:2013pza}. However, one can equivalently use a ten-dimensional perspective, and write the Laplace operator as
\begin{equation}
\label{nablasq_E}
\nabla_{\rm e}^{\, 2} \chi = \frac{1}{\sqrt{G}} \partial_A ( \sqrt{G} G^{AB} \partial_B \chi ) \,,
\end{equation}
where now $G$ and $G^{AB}$ are respectively the determinant and the inverse matrix
 of the ten by ten matrix $G_{AB}$ with $A,B=0,1,...,9$. 

%%%%%%%%%%%%%%%%%%%%%%%%%%%%%%%%%%%%%%%%%
\subsection{Lorentzian signature formulation}

As we assume stationary backgrounds there is no obstruction in Wick rotating the thermal scalar equation of motion to a Lorentzian signature. We write the 
Lorentzian signature metric as
\begin{equation}
ds^2 = g_{\mu\nu} dx^\mu dx^\nu \,,
\end{equation}
with space-time indices $\mu,\nu=0,1,...,9$ in coordinates for which the stationary symmetry is manifest. 
From the Euclidean section metric \eqref{metric_eucl} a metric of this form can be obtained by the Wick rotation $x^0 \equiv t=- i t_{\rm E}$. 
The EOM is now
\begin{equation}
\label{chi_eq}
\nabla^2 \chi = m^2 \chi \,,
\end{equation}
with Laplace operator 
\begin{equation}
\label{nablasq_L}
\nabla^2 \chi = \frac{1}{\sqrt{-g}} \partial_\mu ( \sqrt{-g} g^{\mu\nu} \partial_\nu \chi ) \,.
\end{equation}

It is useful for writing a general version of the thermal scalar formalism to introduce a vector field $V$ to parametrize the flow of time, instead of an explicit time coordinate. 
For coordinate systems in which $t$ is a time-coordinate we have
\begin{equation}
V = \frac{\partial}{\partial t} \,.
\end{equation}
We write now $m^2$ as 
\begin{equation}
\label{m_square_L}
m^2 =- \frac{2}{l_s^2}+ \frac{g_{\mu\nu} \tau^\mu \tau^\nu}{l_s^4}  \,,
\end{equation}
where we introduced the vector field
\begin{equation}
\label{def_tau}
\tau = -  \frac{i \beta}{2\pi} V \,.
\end{equation}
The part $-i V$ can be thought as the Euclidean time flow. We have packaged this together with the inverse temperature $\beta$ to make the point that any dependence of $\beta$ should happen through the vector field $\tau$. This will be important below.
In fact, with $V$ marking the flow of time, a rescaling of time corresponds to a rescaling $V' = \tfrac{1}{\kappa} V$. This should be joined with the rescaling $\beta' = \kappa \beta$, which means $\beta' V' = \beta V$. 

Regarding dimensions, we take the coordinates $x^\mu$ to have dimension length while $g_{\mu\nu}$ and $V^\mu$ are dimensionless. Hence 
\begin{equation}
g_{\mu\nu} \tau^\mu \tau^\nu = - \frac{\beta^2}{4\pi^2} g_{\mu\nu} V^\mu V^\nu \,,
\end{equation}
has dimension length squared, since $\beta$ has dimension of length.

%%%%%%%%%%%%%%%%%%%%%%%%%%%%%%%%%%%%%%%%%
\subsection{Curvature correction for Ramond-Ramond flux}
\label{sec:thermal_scalar_curv}

On general grounds, one expects there should exists an effective description for the thermal scalar that can provide the correct near-Hagedorn dynamics for type II string theory backgrounds. 
However, we shall see explicitly in Sections \ref{sec:ppwave} and \ref{sec:ads} that
the thermal scalar EOM, given by Eqs.~\eqref{chi_eq}, \eqref{nablasq_L} and \eqref{m_square_L},
cannot reproduce the Hagedorn temperature of type II string theory for backgrounds with Ramond-Ramond flux.
Indeed, it fails at the first order in which the curvature of the target space background becomes relevant. 
This is despite the fact that the backgrounds of Sections \ref{sec:ppwave} and \ref{sec:ads} are protected as string theory target space geometries against curvature corrections.

The purpose of this section is to investigate the possible correction terms that one can add to the thermal scalar EOM at lowest order in the curvature of the backgrounds.
Thus, we consider terms that are first order in the Riemann curvature tensor $R_{\mu\nu\rho\sigma}$. Such terms would become zero for flat space, and thus ensure that we still obtain the correct flat space Hagedorn temperature.
Note also that the dimension of both sides of Eq.~\eqref{chi_eq} is $1/\mbox{length}^2$, thus any new term should have this dimension as well. 

On the RHS of Eq.~\eqref{chi_eq} we have $m^2$ given by \eqref{m_square_L}.
What terms can one add that are first order in  $R_{\mu\nu\rho\sigma}$?
The only tensors we can use to contract are $g^{\mu\nu}$ and $\tau^\mu$. As explained above, the vector field $V^\mu$ can only enter in the combination $\beta V^\mu$. Since $R_{\mu\nu\rho\sigma} \tau^\mu \tau^\nu \tau^\rho \tau^\sigma = 0$ this leaves us with the Ricci scalar $R$ and $R_{\mu\nu} \tau^\mu\tau^\nu$. Regarding $R$, all the backgrounds supported by Ramond-Ramond flux that we consider in this paper has zero Ricci scalar 
\begin{equation}
R=0 \,.
\end{equation}
Hence we cannot probe a coupling of this kind. 
This leaves us with the term $l_s^{-2} R_{\mu\nu} \tau^\mu\tau^\nu$ where the string length is introduced to fit the dimension of $m^2$. 
Thus, we consider the curvature correction
\begin{equation}
\label{m_square_curv}
m^2 =- \frac{2}{l_s^2}+ \frac{g_{\mu\nu} \tau^\mu \tau^\nu}{l_s^4} + C \frac{  R_{\mu\nu} \tau^\mu\tau^\nu}{l_s^2} \,,
\end{equation}
where $C$ is an undetermined coefficient. At this order in the curvature, one could also include terms of the form $R_{\mu\nu} \tau^\mu\tau^\nu (g_{\mu\nu}\tau^\mu \tau^\nu)^n$. However, as we shall see in Section \ref{sec:ppwave}, such terms are clearly excluded by comparing with Hagedorn temperatures of type II string theory on pp-wave backgrounds. 

On the LHS of Eq.~\eqref{chi_eq} we have the Laplace operator \eqref{nablasq_L}. 
Here the curvature can couple to the derivatives of $\chi$, {\sl i.e.} $\nabla_\mu \nabla_\nu \chi$.
A possible term is thus $R^{\mu\nu} \nabla_\mu \nabla_\nu \chi$. Instead $R\, g^{\mu\nu} \nabla_\mu \nabla_\nu \chi$ is not relevant as $R=0$ for the backgrounds supported by Ramond-Ramond flux in this paper. One could also entertain a term of the form $\tau^\rho \tau^\sigma R^{\mu}{}_\rho{}^\nu{}_\sigma \nabla_\mu \nabla_\nu \chi$. 
However, this would mix the Laplacian term with the thermal circle which we assume is not a possibility.
Thus, we conclude that the possible curvature correction to the EOM for $\chi$ is
\begin{equation}
\label{chi_eq_curv}
( g^{\mu\nu} - \tilde{C}\, l_s^2 R^{\mu\nu} ) \nabla_\mu \nabla_\nu \chi = m^2 \,,
\end{equation}
to first order in the curvature,
with $m^2$ given by \eqref{m_square_curv} and where $\tilde{C}$ is an  undetermined coefficient.

In Sections \ref{sec:ppwave} and \ref{sec:ads} we find that $C$ and $\tilde{C}$ are fixed to be
\begin{equation}
\label{theCs}
C = \tilde{C} = 2 \log 2 \,.
\end{equation}
We shall see in Sections \ref{sec:ppwave} and \ref{sec:ads} that the value of $C$ is an exact statement, while the one for $\tilde{C}$ fits with current numerical results and conjectures.

In addition to terms coupling to the curvature tensor, as introduced above, one could also have contemplated terms which are quadratic in the Ramond-Ramond field strength, {\sl i.e} a coupling to $(F^{(n)})_{\mu \rho_1 \cdots \rho_{n-1}} (F^{(n)})_{\nu}{}^{\rho_1 \cdots \rho_{n-1}}$ for an $n$-form field strength. However, as we shall comment on below in Section \ref{sec:thermal_scalar_KR}, such terms would not be able to occur independently of $R_{\mu\nu}$ due to the expectation that all correction terms which are first order in the curvature should vanish when no Ramond-Ramond fluxes are present. 

Finally, we note that only corrections with even powers of $l_s$ are expected to occur, since $n$ loops in the sigma-model description enters with coefficient $l_s^{2n}$ relative to the tree-level, which for $m^2$ is the $l_s^{-4} g_{\mu\nu}\tau^\mu\tau^\nu$ term. Therefore, the next correction beyond the first-order curvature terms should be order $l_s^4$ relative to tree-level.

%%%%%%%%%%%%%%%%%%%%%%%%%%%%%%%%%%%%%%%%%
\subsection{Curvature correction by field redefinition of metric}

Consider the curvature corrected mass square given by Eq.~\eqref{m_square_curv}. One can alternatively obtain this from an uncorrected formula for the mass square
\begin{equation}
\label{m_square_tilde}
\tilde{m}^2 =- \frac{2}{l_s^2}+ \frac{\tilde{g}_{\mu\nu} \tau^\mu \tau^\nu}{l_s^4}  \,,
\end{equation}
by a field redefinition of the metric 
\begin{equation}
\label{field_redef}
\tilde{g}_{\mu\nu} = g_{\mu\nu} + C \, l_s^2 R_{\mu\nu}
\end{equation}
How about the correction to the Laplacian \eqref{chi_eq_curv}? 
Assuming \eqref{field_redef}, the inverse metric is
\begin{equation}
\label{field_redef_inv}
\tilde{g}^{\mu\nu} = g^{\mu\nu} - C \, l_s^2 R^{\mu\nu}
\end{equation}
up to higher order corrections in the curvature. Thus,
\begin{equation}
\tilde{\nabla}^2 \chi = \tilde{g}^{\mu\nu} \tilde{\nabla}_\mu \tilde{\nabla}_\nu \chi
=  ( g^{\mu\nu} - C\, l_s^2 R^{\mu\nu} ) \nabla_\mu \nabla_\nu \chi  - g^{\mu\nu} \delta \Gamma^\rho_{\mu\nu} \partial_\rho \chi 
\end{equation}
up to second order corrections in the curvature, where $\delta \Gamma^\rho_{\mu\nu}$ is the correction to the Christoffel Symbol due to the field redefinition \eqref{field_redef}.
We calculate
\begin{equation}
g^{\mu\nu} \delta \Gamma^\rho_{\mu\nu} 
= \frac{1}{2}  g^{\rho\sigma} ( 2 \nabla^\mu \delta g_{\mu\sigma} - \nabla_\sigma g^{\mu\nu} \delta g_{\mu\nu} ) 
= \frac{C\, l_s^2}{2} ( 2 \nabla^\mu R_{\mu\sigma} - \nabla_\sigma R ) =0
\end{equation}
where the last equation follows from the Bianchi identity. Thus, we have found that upon performing the field redefinition \eqref{field_redef},
the uncorrected thermal scalar EOM 
\begin{equation}
\tilde{\nabla}^2 \chi = \tilde{m}^2\chi \,,
\end{equation}
with mass square \eqref{m_square_tilde},
becomes equivalent to the curvature corrected
thermal scalar EOM, given by Eqs.~\eqref{m_square_curv} and \eqref{chi_eq_curv}, 
provided $C=\tilde{C}$. As mentioned above, we find that $C$ and $\tilde{C}$ are fixed to be \eqref{theCs} in Sections \ref{sec:ppwave} and \ref{sec:ads}.
Thus, one can think of the curvature correction to the thermal scalar equation as a field redefinition of the metric.

%%%%%%%%%%%%%%%%%%%%%%%%%%%%%%%%%%%%%%%%%
\subsection{Turning on Kalb-Ramond flux}
\label{sec:thermal_scalar_KR}

Above, we have considered only backgrounds with Ramond-Ramond flux turned on. 
One can extend the thermal scalar formalism to include a non-zero Kalb-Ramond field flux
\begin{equation}
H_{\mu\nu\rho} = (dB)_{\mu\nu\rho} = \partial_\mu B_{\nu\rho} +\partial_\rho B_{\mu\nu} +\partial_\nu B_{\rho\mu} \,,
\end{equation}
as well. If there is no Ramond-Ramond flux present, we have the thermal scalar EOM given by Eqs.~\eqref{chi_eq} and \eqref{nablasq_L} but with $m^2$ modified as \cite{Mertens:2015ola}
\begin{equation}
\label{m_square_B}
m^2 = - \frac{2}{l_s^2}  + \frac{g_{\mu\nu} \tau^\mu \tau^\nu}{ l_s^4} + \frac{g^{\mu\nu}  B_{\mu\rho}  B_{\nu\sigma}\tau^\rho\tau^\sigma}{l_s^4}  \,.
\end{equation}

One expects that the thermal scalar EOM given by Eqs.~\eqref{chi_eq}, \eqref{nablasq_L} and \eqref{m_square_B} describes all general type II string theory backgrounds with constant dilaton, vanishing string coupling and no Ramond-Ramond fields, at least up to first-order corrections in the curvature. Arguments for this can for example by found in \cite{Mertens:2015ola}, and recently it was seen that the Hagedorn temperature can be correctly computed on Kalb-Ramond supported AdS$_3$ geometries using the thermal scalar EOM given by Eqs.~\eqref{chi_eq}, \eqref{nablasq_L} and \eqref{m_square_B} \cite{Urbach:2023npi}.
Below in Section \ref{sec:ppwave_KR} we provide another example of this expectation with a Kalb-Ramond supported pp-wave background. 

The expectation that no there are no corrections to the thermal scalar EOM at first order in the curvature, when no Ramond-Ramond fluxes are present, puts a constraint on how first-order curvature terms can appear, even when turning on Ramond-Ramond fluxes. Indeed, the condition in type II supergravity that there are no Ramond-Ramond fluxes is 
\begin{equation}
R_{\mu\nu} - \frac{1}{4} H_{\mu\rho\sigma} H_{\nu} {}^{\rho\sigma} =0 \,,
\end{equation}
when assuming a constant dilaton. In order to have a first-order curvature term that vanish when there is no Ramond-Ramond flux, such a term needs to couple to the tensor $R_{\mu\nu} - \frac{1}{4} H_{\mu\rho\sigma} H_{\nu} {}^{\rho\sigma}$, or a contraction of this. This also means that one cannot introduce couplings to $n$-form Ramond-Ramond field strengths through $((F^{(n)})^2)_{\mu\nu} = (F^{(n)})_{\mu \rho_1 \cdots \rho_{n-1}} (F^{(n)})_{\nu}{}^{\rho_1 \cdots \rho_{n-1}}$ which are independent of the tensor $R_{\mu\nu} - \frac{1}{4} H_{\mu\rho\sigma} H_{\nu} {}^{\rho\sigma}$, thus explaining why we did not need to take such terms into account above in Section \ref{sec:thermal_scalar_curv}.%
\footnote{In addition, note that the coupling to Ramond-Ramond field strengths should be invariant under T-duality, thus constraining the form of the sum over the couplings to $((F^{(n)})^2)_{\mu\nu}$ to be proportional to $R_{\mu\nu} - \frac{1}{4} H_{\mu\rho\sigma} H_{\nu} {}^{\rho\sigma}$.}

Considering the curvature corrections \eqref{m_square_curv} and \eqref{chi_eq_curv} for zero Kalb-Ramond flux, it is evident how to generalize this to have a coupling to the tensor $R_{\mu\nu} - \frac{1}{4} H_{\mu\rho\sigma} H_{\nu} {}^{\rho\sigma}$. 
We conjecture that the curvature corrections in the presence of a Kalb-Ramond flux are modified as
\begin{equation}
\label{chi_eq_curv_B}
\Big( g^{\mu\nu} - \tilde{C}\, l_s^2 ( R^{\mu\nu} -\tfrac{1}{4} H^{\mu}{}_{\rho\sigma} H^{\nu\rho\sigma} )\Big) \nabla_\mu \nabla_\nu \chi = m^2 \,,
\end{equation}
with $m^2$ given by
\begin{equation}
\label{m_square_curv_B}
m^2 = - \frac{2}{l_s^2}  + \frac{g_{\mu\nu} \tau^\mu \tau^\nu}{ l_s^4} + \frac{g^{\mu\nu}  B_{\mu\rho}  B_{\nu\sigma}\tau^\rho\tau^\sigma}{l_s^4} 
+ C \frac{  (R_{\mu\nu} -\tfrac{1}{4} H_{\mu\rho\sigma} H_{\nu}{}^{\rho\sigma} ) \tau^\mu\tau^\nu}{l_s^2} \,.
\end{equation}
In Section \ref{sec:ppwave_KR} we consider the above thermal scalar EOM in the case of a pp-wave background with both Ramond-Ramond and Kalb-Ramond fluxes present, and find agreement with the Hagedorn temperature computed by the spectrum when assuming \eqref{theCs}.

%%%%%%%%%%%%%%%%%%%%%%%%%%%%%%%%%%%%%%%%%
\section{Hagedorn temperature on pp-wave backgrounds}
\label{sec:ppwave}

We show in this section that the thermal scalar formalism of Section \ref{sec:thermal_scalar} can be used on pp-wave backgrounds of type II string theory.

%%%%%%%%%%%%%%%%%%%%%%%%%%%%%%%%%%%%%%%%%
\subsection{Ramond-Ramond flux supported pp-waves}
\label{sec:ppwave_RR}

We consider three particular pp-wave backgrounds of type II supergravity, supported by Ramond-Ramond flux. All of them are described by the metric
\begin{equation}
\label{metric_ppwave}
ds^2 = - 2 dx^+ dx^- - f^2 \left( a_1^{\, 2} \sum_{I=1}^4 x_I^{\, 2} + a_2^{\, 2} \sum_{I=5}^8 x_I^{\, 2} \right) (dx^+)^2 + \sum_{I=1}^8 dx_I^{\, 2} \,.
\end{equation}
The Ramond-Ramond flux in the three cases are given by \cite{Blau:2001ne,Berenstein:2002jq,Sugiyama:2002tf,Hyun:2002wu,Russo:2002rq}
\begin{equation}
\label{RRflux_ppwave}
\begin{array}{ll}
\mbox{5-form flux:} &\ \ F_{+1234} = F_{+5678} = 2f \,,
\\[2mm]
\mbox{2 and 4-form flux:} &\ \  F_{+123} = f \spa F_{+4} = \frac{f}{3} \,,
\\[2mm]
\mbox{3-form flux:} &\ \  F_{+12}=F_{+34}= 2 f  \,,
\end{array}
\end{equation}
where $F_{\mu_1 \cdots \mu_n}$ is the $n$-form Ramond-Ramond field strength. 
The constants $a_1$ and $a_2$ in the metric \eqref{metric_ppwave} are given as
\begin{equation}
\label{the_ai}
\begin{array}{ll}
\mbox{5-form flux:} &\ \ a_1=a_2=1 \,,
\\[2mm]
\mbox{2 and 4-form flux:} &\ \  a_1=\frac{1}{3} \spa a_2 = \frac{1}{6} \,,
\\[2mm]
\mbox{3-form flux:} &\ \  a_1=1 \spa a_2=0 \,.
\end{array}
\end{equation}
These three pp-waves can be obtained from Penrose limits of the AdS backgrounds that we consider below in Section \ref{sec:ads}. We have
\begin{equation}
\begin{array}{ll}
\mbox{5-form flux:} &\ \ \mbox{Penrose limit of AdS$_5 \times S^5$} \,,
\\[2mm]
\mbox{2 and 4-form flux:} &\ \  \mbox{Penrose limit of AdS$_4 \times \C \mbox{P}^3$}\,,
\\[2mm]
\mbox{3-form flux:} &\ \  \mbox{Penrose limit of AdS$_3 \times S^3 \times T^4$}\,.
\end{array}
\end{equation}
The pp-wave backgrounds above are known to be exact backgrounds of string theory with respect to $\alpha'$ corrections \cite{Horowitz:1989bv,Kallosh:1998qs}. 

%%%%%%%%%%%%%%%%%%%%%%%%%%%%%%%%%%%%%%%%%
\subsection{Hagedorn temperature from string spectrum}
\label{sec:ppwave_TH_partition}

For the three cases of pp-waves given by Eqs.~\eqref{metric_ppwave}, \eqref{RRflux_ppwave} and \eqref{the_ai} the spectrum of a closed string in type II string theory from lightcone gauge quantization is given by \cite{Metsaev:2001bj,Berenstein:2002jq,Metsaev:2002re,Sugiyama:2002tf,Hyun:2002wu}
\begin{equation}
\label{ppwave_spec}
p^- = \frac{1}{p^+} \sum_{m\in \Z} \left[ \sum_{I=1}^4 \sqrt{m^2 + a_1^{\, 2} \mu^2 } \Big( N^{(I)}_m + F^{(I)}_m \Big) +\sum_{I=5}^8 \sqrt{m^2 + a_2^{\, 2} \mu^2 } \Big( N^{(I)}_m + F^{(I)}_m \Big) \right] \,.
\end{equation}
Here $N^{(I)}_m$ and $F^{(I)}_m$ are the bosonic and fermionic number operators, respectively, counting the oscillators at level $m$, with $I=1,...,8$. The level-matching condition is
\begin{equation}
\label{ppwave_spec2}
 \sum_{I=1}^8 \sum_{m\in \Z} m \left(  N_{m}^{(I)} +  F_{m}^{(I)} \right) =0 \,,
\end{equation}
and the constant $\mu$ in \eqref{ppwave_spec} is defined by
\begin{equation}
\label{ppwave_spec3}
\mu = l_s^2 p^+ f \,.
\end{equation}

The Hagedorn temperature for string theory in the pp-wave backgrounds given by Eqs.~\eqref{metric_ppwave}, \eqref{RRflux_ppwave} and \eqref{the_ai} corresponds to the temperature at which the expression for the free energy of a gas of non-interacting closed strings, {\sl i.e.}~strings that do not split or join, starts to become ill-defined. This is directly tied to the exponential growth of closed string states when close to the Hagedorn temperature. 
To obtain the Hagedorn temperature, we first remark that we choose here to measure temperature with respect to the time $t$, related to the lightcone directions as
\begin{equation}
\label{time_choice}
x^\pm = \frac{1}{\sqrt{2}} ( t \pm x_9 ) \,,
\end{equation}
where $x_9$ is a spatial direction. This means the free energy can be written as
\begin{equation}
\label{free_energy}
F = - \sum_{n,\rm{odd}}^\infty \frac{1}{n\beta} \mbox{Tr} \left( e^{-n \beta p^0}\right)
= - \sum_{n,\rm{odd}}^\infty \frac{1}{n\beta} \mbox{Tr} \left( e^{-\frac{n\beta}{\sqrt{2}} (p^++p^-)}\right) \,,
\end{equation}
in terms of the inverse temperature $\beta=1/T$ and
where we have ignored an extra term due to excitations with $p^-=0$ that in any case do not contribute to Hagedorn behavior. Since the Hagedorn temperature arise from the $n=1$ contribution, corresponding to single-string dominance, we focus on this contribution. One finds 
\begin{equation}
\label{free_energy_n1}
F|_{n=1} \propto \int_0^\infty \frac{d\tau_2}{\tau_2^2}  \int_{-\tfrac{1}{2}}^{\tfrac{1}{2}} d\tau_1  e^{- \frac{\beta^2}{4\pi l_s^2 \tau_2} + Y (\beta,\tau_1,\tau_2)} \,,
\end{equation}
\begin{equation}
Y (\beta,\tau_1,\tau_2) 
= 4 \sum_{i=1}^2 \sum_{m\in \Z}  \log \frac{1+e^{-2\pi \tau_2 \sqrt{m^2 + a_i^2 \mu^2 } + 2\pi i \tau_1 m}}{1-e^{-2\pi \tau_2 \sqrt{m^2 + a_i^2 \mu^2 } + 2\pi i \tau_1 m}}  \,.
\end{equation}
Here $\tau_1$ is introduced as Lagrange multiplier to account for level matching and we defined
\begin{equation}
\tau_2 = \frac{\beta}{2\sqrt{2} \pi l_s^2 p^+} \,.
\end{equation}
The Hagedorn temperature corresponds to a divergence in the free energy for $\tau_1,\tau_2 \rightarrow 0$. Setting first $\tau_1=0$ we have
\begin{equation}
Y (\beta,\tau_1,\tau_2) 
= 4 \sum_{i=1}^2\sum_{m\in \Z}\sum_{p=1}^\infty \frac{1-(-1)^p}{p} e^{-2\pi p \tau_2 \sqrt{m^2 + a_i^2 \mu^2 }} \,.
\end{equation}
Using methods explained in \cite{Grignani:2003cs} one can obtain the following estimate for small $\tau_2$
\begin{equation}
Y (\beta,\tau_1,\tau_2) 
\simeq  \frac{4}{\tau_2} \Big[H(a_ 1\tfrac{\beta f}{\sqrt{2}} )+ H(a_ 2\tfrac{\beta f}{\sqrt{2}} )\Big] \,,
\end{equation}
where we defined
\begin{equation}
H(x) =  \frac{x}{\pi} \sum_{p=1}^\infty \frac{1-(-1)^p}{p} K_1(xp) \,.
\end{equation}
Thus, from \eqref{free_energy_n1} we see the Hagedorn temperature is determined by the equation
\begin{equation}
\label{ppwave_TH_eq1}
\frac{\beta^2}{4\pi l_s^2} = 4 H(a_ 1\tfrac{\beta f}{\sqrt{2}} )+4 H(a_ 2\tfrac{\beta f}{\sqrt{2}} ) \,,
\end{equation}
since the exponential in \eqref{free_energy_n1} diverges for $\tau_2\rightarrow 0$ when the temperature exceeds the Hagedorn temperature. The formula \eqref{ppwave_TH_eq1} reduces to the explicit Hagedorn temperature computed in \cite{PandoZayas:2002hh,Greene:2002cd,Sugawara:2002rs,Brower:2002zx,Grignani:2003cs,Hyun:2003ks} for the three cases \eqref{the_ai}.
Following \cite{Grignani:2003cs} one can reformulate $H(x)$ as the infinite series
\begin{equation}
H(x) =\frac{\pi}{4} - \frac{1}{2} x + \frac{\log 2}{2\pi} x^2  
- \sum_{k=2}^\infty \frac{(-1)^k}{k!}  \frac{\Gamma(k-\tfrac{1}{2})}{\sqrt{\pi}} \zeta(2k-1)\frac{2^{2k-2}-1}{(2\pi)^{2k-1}}   x ^{2k} \,.
\end{equation}
Thus, for small $x$ we have
\begin{equation}
H(x) =\frac{\pi}{4} - \frac{1}{2} x + \frac{\log 2}{2\pi} x^2  -  \frac{3\zeta(3)}{32\pi^3}  x^4
+\frac{15\zeta(5)}{256 \pi^5} x^6 - \frac{315 \zeta(7)}{8192\pi^7} x^8
+ \CO (x^{10}) \,.
\end{equation}
Therefore, the Hagedorn temperature equation is
\begin{eqnarray}
\label{ppwave_TH_eq2}
\frac{\beta^2}{4\pi l_s^2} &=& 2\pi - \sqrt{2} (a_1+a_2) \beta f + \frac{\log 2}{\pi} (a_1^{\, 2}+a_2^{\, 2}) \beta^2 f^2 - \frac{3\zeta(3)}{32\pi^3} (a_1^{\, 4}+a_2^{\, 4}) \beta^4 f^4
\nn \\ && +\frac{15\zeta(5)}{2^9 \pi^5}  (a_1^{\, 6} + a_2^{\, 6}) \beta^6 f^6 - \frac{315 \zeta(7)}{2^{15}\pi^7} (a_1^{\, 8} + a_2^{\, 8}) \beta^8 f^8 + \CO((\beta f)^{10}) \,,
\end{eqnarray}
for $\beta f \ll 1$. Inserting the $a_1$ and $a_2$ coefficients \eqref{the_ai} for the three pp-waves one obtains the Hagedorn temperature expanded in powers of $\beta f$ for each case.

%%%%%%%%%%%%%%%%%%%%%%%%%%%%%%%%%%%%%%%%%
\subsection{Hagedorn temperature from thermal scalar}
\label{sec:ppwave_thermal_scalar}

Using the thermal scalar introduced in Section \ref{sec:thermal_scalar}, we now compute the Hagedorn temperature of the three pp-wave backgrounds of type II string theory given by Eqs.~\eqref{metric_ppwave}, \eqref{RRflux_ppwave} and \eqref{the_ai}.
These three pp-wave backgrounds are known to be exact backgrounds of string theory with respect to $\alpha'$ corrections \cite{Horowitz:1989bv,Kallosh:1998qs} hence on general grounds one would expect it to be possible to obtain the exact Hagedorn temperature from the thermal scalar EOM.

To begin, we should identify the flow-of-time vector field $V$ in the geometry \eqref{metric_ppwave}. By the choice of time \eqref{time_choice}, we have
\begin{equation}
V = \frac{\partial}{\partial t} = \frac{1}{\sqrt{2}} \left( \frac{\partial}{\partial x^+} +   \frac{\partial}{\partial x^-} \right) 
\end{equation}
Using \eqref{def_tau} we get
\begin{equation}
\label{ppwave_tau}
\tau = - \frac{i\beta}{2\sqrt{2}\pi} \left( \frac{\partial}{\partial x^+} +   \frac{\partial}{\partial x^-} \right) 
\end{equation}

To begin, consider first the leading order thermal scalar EOM given by Eqs.~\eqref{chi_eq}, \eqref{nablasq_L} and \eqref{m_square_L}.
From the metric \eqref{metric_ppwave} we find that $\sqrt{-g}=1$. Moreover, the non-zero components of the inverse metric are $g^{+-}=-1$, $g^{IJ}=\delta^{IJ}$ and $g^{--} = - g_{++}$.
This means the Laplace operator is simply $\nabla^2 \chi= \sum_{I=1}^8 \partial_I^2 \chi$ where we assume $\chi$ only depends on the transverse directions $x_I$ as we do not expect the isometries along $x^\pm$ to be broken by $\chi$. Thus, the thermal scalar EOM is
\begin{equation}
\label{ddchi_ppwave}
\sum_{I=1}^8 \partial_I^2 \chi = m^2 \chi
\end{equation}
It is now straightforward to compute the leading order $m^2$ from \eqref{m_square_L} using \eqref{ppwave_tau}
\begin{equation}
\label{ppwave_leading_m2}
m^2 = - \frac{2}{l_s^2}  + \frac{\beta^2}{4\pi^2 l_s^4} \left[  1 + \frac{1}{2} f^2 \left( a_1^2 \sum_{I=1}^4 x_I^2 + a_2^2 \sum_{I=5}^8 x_I^2 \right) \right]
\end{equation}

Introduce now the ansatz for $\chi$
\begin{equation}
\label{chi_ansatz_ppwave}
\chi = A \exp \left( - b_1  \sum_{I=1}^4 x_I^2 - b_2 \sum_{I=5}^8 x_I^2  \right)
\end{equation}
for arbitrary constants $A$, $b_1$ and $b_2$.
We compute
\begin{equation}
\label{ddchi_ansatz_ppwave}
\sum_{I=1}^8 \partial_I^2 \chi = \left[ - 8 (b_1+b_2) + 4 b_1^2 \sum_{I=1}^4 x_I^2 + 4 b_2^2 \sum_{I=5}^8 x_I^2 \right] \chi
\end{equation}
Equating \eqref{ddchi_ansatz_ppwave} with \eqref{ddchi_ppwave} and \eqref{ppwave_leading_m2} we obtain 
\begin{equation}
\label{the_bi}
b_i = \frac{\beta f}{4\sqrt{2}\pi l_s^2} a_i
\end{equation}
as well as
\begin{equation}
-8 (b_1+b_2) = - \frac{2}{l_s^2}  + \frac{\beta^2}{4\pi^2 l_s^4} 
\end{equation}
from which we infer
\begin{equation}
\frac{\beta^2}{4\pi l_s^2} =  2\pi - \sqrt{2} (a_1+a_2) \beta f 
\end{equation}
Comparing this result to the expanded equation for the Hagedorn temperature \eqref{ppwave_TH_eq2} from the type II string spectrum, we see that while the LHS is identical, the RHS reproduces the first two terms of Eq.~\eqref{ppwave_TH_eq2}. Our interpretation of this is that the leading order thermal scalar EOM, given by Eqs.~\eqref{chi_eq}, \eqref{nablasq_L} and \eqref{m_square_L}, is corrected by terms that probes the curvature of the metric \eqref{metric_ppwave}, precisely as advocated in Section \ref{sec:thermal_scalar_curv}. 

Consider therefore the Riemann curvature tensor $R_{\mu\nu\rho\sigma}$ of \eqref{metric_ppwave}. The non-zero components of  $R_{\mu\nu\rho\sigma}$ are
\begin{equation}
\label{Riemann_ppwave}
R_{+I+J} = R_{I+J+} = - R_{I++J}= - R_{+IJ+} = \left\{ \begin{array}{l} a_1^2 f^2\delta_{IJ} \ \ \mbox{for} \ \ I,J = 1,2,3,4 \\  a_2^2 f^2\delta_{IJ} \ \ \mbox{for} \ \ I,J = 5,6,7,8 \end{array} \right.
\end{equation}
The only non-zero component of $R_{\mu\nu}$ is therefore
\begin{equation}
\label{Ricci_ppwave}
R_{++} = 4 (a_1^{\, 2}+a_2^{\, 2} ) f^2
\end{equation}
whereas the Ricci scalar $R=0$.

Consider now the curvature corrected thermal scalar EOM given by Eqs.~\eqref{m_square_curv} and \eqref{chi_eq_curv}. Notice first that since $R^{IJ} = 0$ the LHS of Eq.~\eqref{chi_eq_curv} is not corrected. Thus, the thermal scalar equation is still given by Eq.~\eqref{ddchi_ppwave}.
Instead for $m^2$ given by Eq.~\eqref{m_square_curv}  we compute
\begin{equation}
\label{ppwave_corr_m2}
m^2 = - \frac{2}{l_s^2}  + \frac{\beta^2}{4\pi^2 l_s^4} \left[  1 + \frac{1}{2} f^2 \left( a_1^2 \sum_{I=1}^4 x_I^2 + a_2^2 \sum_{I=5}^8 x_I^2 \right) \right] - C (a_1^2+a_2^2) \frac{\beta^2f^2}{2\pi^2 l_s^2} 
\end{equation}
using \eqref{ppwave_tau} and \eqref{Ricci_ppwave}.
Using again the ansatz \eqref{chi_ansatz_ppwave}, we can equate 
\eqref{ddchi_ansatz_ppwave} with \eqref{ddchi_ppwave} and \eqref{ppwave_corr_m2} to obtain
\eqref{the_bi} along with
\begin{equation}
-8 (b_1+b_2) = - \frac{2}{l_s^2}  + \frac{\beta^2}{4\pi^2 l_s^4} - C (a_1^2+a_2^2) \frac{\beta^2f^2}{2\pi^2 l_s^2} 
\end{equation}
The Hagedorn temperature should thus obey 
\begin{equation}
\label{TH_ppwave_curv}
\frac{\beta^2}{4\pi l_s^2} =  2\pi - \sqrt{2} (a_1+a_2) \beta f + \frac{C}{2\pi} (a_1^2+a_2^2) \beta^2f^2
\end{equation}
Comparing again to the expanded Hagedorn temperature \eqref{ppwave_TH_eq2} for small $\beta f$, we see that the curvature corrected expression \eqref{TH_ppwave_curv} matches  \eqref{ppwave_TH_eq2}, at order $\beta^2 f^2$, provided we identify
\begin{equation}
\label{theC}
C = 2 \log 2
\end{equation}
With this choice, we have matched the first contribution to the Hagedorn temperature \eqref{ppwave_TH_eq2} that arise from the curvature of the pp-wave background \eqref{metric_ppwave}. Thus, we have obtained this contribution from the curvature corrected thermal scalar EOM, given by Eqs.~\eqref{m_square_curv} and \eqref{chi_eq_curv},
 for the three Ramond-Ramond flux supported exact pp-wave backgrounds of type II string theory given by Eqs.~\eqref{metric_ppwave}, \eqref{RRflux_ppwave} and \eqref{the_ai}.

%%%%%%%%%%%%%%%%%%%%%%%%%%%%%%%%%%%%%%%%%
\subsection{Adding a chemical potential}
\label{sec:ppwave_chempot}

To further examine the ability of the thermal scalar to predict the Hagedorn temperature, we consider here adding a chemical potential. 
More specifically, we consider the free energy 
\begin{equation}
\label{free_energy_omega}
F(\beta,\omega,f) = - \sum_{n,\rm{odd}}^\infty \frac{1}{n\beta} \mbox{Tr} \left( e^{-n \beta( p^0+\omega p^9)}\right)
\end{equation}
where we introduced a chemical potential $\omega$ for the momentum $p^9$. Using \eqref{time_choice} we can write this as
\begin{equation}
\label{free_energy_omega2}
F(\beta,\omega,f) = - \sum_{n,\rm{odd}}^\infty \frac{1}{n\beta} \mbox{Tr} \left( e^{-\frac{n\beta}{\sqrt{2}} ((1-\omega) p^++(1+\omega) p^-)}\right)
\end{equation}
The string spectrum given by Eqs.~\eqref{ppwave_spec}-\eqref{ppwave_spec3} has the symmetry
\begin{equation}
p^+ \rightarrow p^+/\Lambda \spa p^- \rightarrow p^-\Lambda \spa f \rightarrow f\Lambda 
\end{equation}
Choosing $\Lambda=\sqrt{1-\omega}/\sqrt{1+\omega}$ one can deduce
\begin{equation}
F(\beta,\omega,f) = F ( \beta \sqrt{1-\omega^2} , 0 , f \sqrt{\tfrac{1-\omega}{1+\omega}} )
\end{equation}
Therefore, the expanded Hagedorn temperature with inverse temperature $\beta$ and chemical potential $\omega$ becomes
\begin{eqnarray}
\label{ppwave_TH_eq_omega}
\frac{\beta^2(1-\omega^2)}{4\pi l_s^2} &=& 2\pi - \sqrt{2} (a_1+a_2) (1-\omega)\beta f + \frac{\log 2}{\pi} (a_1^{\, 2}+a_2^{\, 2})(1-\omega)^2 \beta^2 f^2 
\nn \\ && 
- \frac{3\zeta(3)}{32\pi^3} (a_1^{\, 4}+a_2^{\, 4}) (1-\omega)^4 \beta^4 f^4+\frac{15\zeta(5)}{2^9 \pi^5}  (a_1^{\, 6} + a_2^{\, 6}) (1-\omega)^6 \beta^6 f^6 
\nn \\[2mm] &&
- \frac{315 \zeta(7)}{2^{15}\pi^7} (a_1^{\, 8} 
+ a_2^{\, 8}) (1-\omega)^8 f^8 \beta^8 + \CO((\beta f)^{10})
\end{eqnarray}

From point of view of the thermal scalar, one can view adding the chemical potential as choosing a different flow-of-time vector field
\begin{equation}
V = \frac{1}{\sqrt{2}} \left( (1-\omega) \frac{\partial}{\partial x^+} +(1+\omega) \frac{\partial}{\partial x^-} \right)
\end{equation}
since this new vector field corresponds to choosing a slicing of time with energy $(1-\omega) p^++(1+\omega) p^-$ as in \eqref{free_energy_omega2}.
It is now a straightforward exercise to repeat the derivation of Section \ref{sec:ppwave_thermal_scalar} with the modified flow-of-time vector field $V$, giving
\begin{equation}
\frac{\beta^2(1-\omega^2)}{4\pi l_s^2} =  2\pi - \sqrt{2} (a_1+a_2) (1-\omega) \beta f + \frac{C}{2\pi} (a_1^2+a_2^2) (1-\omega)^2\beta^2f^2
\end{equation}
This reproduces indeed the expanded Hagedorn temperature \eqref{ppwave_TH_eq_omega} up to and including the $\beta^2 f^2$ term. This confirms that the terms in \eqref{m_square_curv} coupling to the vector field $\tau$ transforms correctly when changing the slicing of time.

%%%%%%%%%%%%%%%%%%%%%%%%%%%%%%%%%%%%%%%%%
\subsection{Ramond-Ramond and Kalb-Ramond flux supported pp-wave}
\label{sec:ppwave_KR}

We consider now a pp-wave background with both Ramond-Ramond and Kalb-Ramond flux, in order to check the thermal scalar EOM proposal of Section \ref{sec:thermal_scalar_KR}. To this end, we employ the following pp-wave background with metric \cite{Russo:2002rq}
\begin{equation}
\label{ppwave_KR}
ds^2 = - 2 dx^+dx^- - f^2 \sum_{I=1}^4 x_I^{\, 2} (dx^+)^2 + \sum_{I=1}^8 dx_I^{\, 2} 
\end{equation}
This is supported by the Kalb-Ramond field
\begin{equation}
B_{+1} = f \cos \alpha \, x^2 \spa B_{+2} = - f \cos \alpha \, x^1 \spa B_{+3} =f \cos \alpha \, x^4 \spa B_{+4} = -f \cos \alpha\, x^3 
\end{equation}
corresponding to the 3-form field strength
\begin{equation}
H_{+12}=H_{+34}= 2 f \cos \alpha 
\end{equation}
In addition we have the Ramond-Ramond 3-form flux
\begin{equation}
F_{+12}=F_{+34}= 2 f \sin \alpha
\end{equation}
In the above solution one can adjust the angle $\alpha$ to interpolate between a pp-wave purely supported by Kalb-Ramond flux for $\alpha=0$ and one purely supported by Ramond-Ramond flux for $\alpha = \pi /2$.

Using lightcone quantization for type IIB string theory, one finds that half of the fermionic and bosonic modes have dispersion relation of flat space $p^- = \tfrac{1}{p^+} m$ and the other half $p^- = \tfrac{1}{p^+} \sqrt{\mu^2 \sin^2 \alpha + ( m + \mu \cos \alpha)^2 }$  \cite{Grignani:2003cs}.
Performing the equivalent calculation as in Section \ref{sec:ppwave_TH_partition}, with time direction given by \eqref{time_choice}, one finds the following condition determining the Hagedorn temperature \cite{Grignani:2003cs}
\begin{equation}
\label{ppwave_TH_KR}
\frac{\beta^2}{4\pi l_s^2} = 4 H(\sin \alpha \tfrac{\beta f}{\sqrt{2}} ) + \pi
\end{equation}
Expanding for $|\sin \alpha| \beta f \ll 1$ this gives
\begin{equation}
\label{expanded_TH_KR}
\frac{\beta^2}{4\pi l_s^2} = 2\pi - \sqrt{2} \sin \alpha\, \beta f + \frac{\log 2}{\pi} \sin^2 \alpha\, \beta^2 f^2 - \frac{3\zeta(3)}{32\pi^3} \sin^4 \alpha\, \beta^4 f^4
+ \cdots
\end{equation}

We now consider this case from the thermal scalar point of view.
We use $\tau$ as given by \eqref{ppwave_tau}.
We compute
\begin{equation}
g^{\mu\nu}  B_{\mu\rho}  B_{\nu\sigma}\tau^\rho\tau^\sigma
= - \frac{\beta^2}{8\pi^2} f^2 \cos^2\alpha \sum_{I=1}^4 x_I^{\, 2}
\spa
(R_{\mu\nu} -\tfrac{1}{4} H_{\mu\rho\sigma} H_{\nu}{}^{\rho\sigma} ) \tau^\mu\tau^\nu
=- \frac{\sin^2 \alpha\,  \beta^2f^2}{2\pi^2} 
\end{equation}
Hence from $m^2$ in Eq.~\eqref{m_square_curv_B} we find
\begin{equation}
\label{ppwave_KR_m2}
m^2 = - \frac{2}{l_s^2}  + \frac{\beta^2}{4\pi^2 l_s^4} \left[  1 + \frac{1}{2} \sin^2 \alpha \, f^2  \sum_{I=1}^4 x_I^2  \right] - C  \frac{\sin^2 \alpha \, \beta^2f^2}{2\pi^2 l_s^2} 
\end{equation}
It is straightforward to see that the RHS side of thermal scalar EOM \eqref{chi_eq_curv_B} is not corrected, hence the EOM is \eqref{ddchi_ppwave} with $m^2$ given by \eqref{ppwave_KR_m2}. Repeating now the procedure from Section \ref{sec:ppwave_thermal_scalar} we find the Hagedorn temperature equation
\begin{equation}
\frac{\beta^2}{4\pi l_s^2} = 2\pi - \sqrt{2} \sin \alpha\, \beta f + \frac{C}{2\pi} \sin^2 \alpha\, \beta^2 f^2 
\end{equation}
which correctly matches the first terms of \eqref{expanded_TH_KR} for $C=2\log 2$, up to and including the $\beta^2 f^2$ term. This is evidence that the thermal scalar EOM of 
Section \ref{sec:thermal_scalar_KR} is correct up to first order in the curvature of the background. 

Consider also the special case of $\alpha=0$ for which the pp-wave geometry \eqref{ppwave_KR} is purely supported by a Kalb-Ramond field. In this case the curvature term vanishes, hence the Hagedorn temperature equation becomes exact, predicting correctly the Hagedorn temperature to be the same as for flat space.

%%%%%%%%%%%%%%%%%%%%%%%%%%%%%%%%%%%%%%%%%
\section{Hagedorn temperatures on AdS backgrounds}
\label{sec:ads}

In this section we show that the thermal scalar EOM with curvature correction can reproduce higher-order terms in the Hagedorn temperature for type II string theory on  AdS$_{d+1}$ with the appropriate product space. 

%%%%%%%%%%%%%%%%%%%%%%%%%%%%%%%%%%%%%%%%%
\subsection{AdS backgrounds and Hagedorn temperature from integrability}
\label{sec:ads_backgr}

We introduce here backgrounds of type II string theory with an AdS part supported by Ramond-Ramond flux. Type II string theory on such backgrounds are holographically dual to certain supersymmetry field theories without gravity. Moreover, they are protected from $\alpha'$ corrections \cite{Kallosh:1998qs}. We exhibit these backgrounds and briefly review the recent results on getting their Hagedorn temperature from integrability.

The first background we consider is AdS$_5\times S^5$ in the global patch, given as
\begin{equation}
\label{ads5metric}
ds^2 = - (1+r^2) dt^2 + \frac{dr^2}{1+r^2} +r^2 d\Omega_3^{\, 2} + d\Omega_5^{\, 2}   \,,
\end{equation}
\begin{equation}
\label{ads5F}
F_{(5)}=2  \Big( r^3 dt \, dr \, d\Omega_3 + d\Omega_5 \Big) \,,
\end{equation}
with no Kalb-Ramond flux and constant dilaton. This background is maximally symmetric with 32 supercharges and is exact, in particular with no $\alpha'$ curvature corrections \cite{Kallosh:1998qs}. 
Notice in Eqs.~\eqref{ads5metric} and \eqref{ads5F} that we have absorbed the radius of AdS$_5$ and $S^5$ into the string tension by rescaling of the fields. This means the rescaled string tension is given by $1/(2\pi l_s^2) = L^2 /(2\pi l_s^{\rm (old)}{}^2)$ where $L$ is the radius of AdS$_5$ and $S^5$ and $l_s^{\rm (old)}$ is the original string length. This makes the rescaled string length $l_s$ dimensionless. For use below we also note that the metric \eqref{ads5metric} has zero Ricci scalar $R=0$.

Type IIB string theory on the global-patch AdS$_5\times S^5$ background is holographically dual to $\CN=4$ super-Yang-Mills theory (SYM) on a three-sphere \cite{Maldacena:1997re,Witten:1998qj,Gubser:1998bc}. We write the rescaled string length as the dimensionless quantity
\begin{equation}
\label{string_length}
l_s = \frac{1}{\sqrt{4\pi g}} 
\end{equation}
where $g$ is defined as 
\begin{equation}
g^2 = \frac{\lambda}{16 \pi^2} 
\end{equation}
with $\lambda$ being the t' Hooft coupling of $\CN=4$ super-Yang-Mills theory.

Amazingly, one can use the holographic duality between type IIB string theory on AdS$_5\times S^5$ and $\CN=4$ SYM to study the Hagedorn temperature \cite{Sundborg:1999ue,Aharony:2003sx}.
The Hagedorn temperature requires a vanishing string coupling. This corresponds to the limit with an infinite number of colors of $\CN=4$ super-Yang-Mills theory in which one has the powerful symmetry of integrability  \cite{Beisert:2010jr,Bombardelli:2016rwb}. It was demonstrated in \cite{Harmark:2017yrv,Harmark:2018red,Harmark:2021qma} that one can use integrability to extract the Hagedorn temperature by solving so-called Quantum Spectral Curve (QSC) equations \cite{Gromov:2013pga,Gromov:2014bva,Gromov:2014caa}. This was used to obtain the perturbative 7-loop expression at weak 't Hooft coupling for the Hagedorn temperature exactly, and to find a numerical curve for the Hagedorn temperature in the interval $0 \leq \sqrt{g} \leq 1.8$ (corresponding to $0 \leq \lambda \leq 1660$). From this, the following asymptotic formula is extracted \cite{Harmark:2018red,Harmark:2021qma}%
\footnote{Here the uncertainties are in the last digits in the first two coefficients while for the third and fourth it is $0.0005$ and $0.005$, respectively.}
\begin{equation}
\label{TH_AdS5_HW}
T_{\rm H} = 0.3989 \sqrt{g} + 0.159 - \frac{0.0087}{\sqrt{g}} + \frac{0.037}{g} + \CO( g^{-3/2} )
\end{equation}
for large $g$. 
In \cite{Ekhammar:2023glu} the QSC computation of the Hagedorn temperature was repeated with higher precision and reaching the value $\sqrt{g}=2.25$ ($\lambda= 4050$) of the coupling. This gave the estimate%
\footnote{Here the error is 1 in the last digit for the first four coefficients and 3 for the last two.}
\begin{equation}
\label{TH_AdS5_M}
T_{\rm H} = 0.39894 \sqrt{g} + 0.15916 - \frac{0.00865}{\sqrt{g}} + \frac{0.0356}{g} 
- \frac{0.008196}{g^{3/2}} - \frac{0.00671}{g^2} 
+ \CO( g^{-5/2} )
\end{equation}
for large $g$.

The second background is AdS$_4\times \C \mbox{P}^3$ with metric
\begin{equation}
\label{metric_ads4}
ds^2 =   - (1+r^2) dt^2 + \frac{dr^2}{1+r^2} +r^2 d\Omega_2^{\, 2} + 4 ds_{\C \rm P^3}^{\, 2} \,,
\end{equation}
where $ds_{\C \rm P^3}^{\, 2}$ is the Fubini-Study metric on $\C \mbox{P}^3$.
This is supported by a four-form Ramond-Ramond flux along AdS$_4$ and a two-form Ramond-Ramond flux proportional to the K\" ahler form on $\C \mbox{P}^3$.
The dilaton is zero and there is no Kalb-Ramond flux.
For use below we note that the metric \eqref{metric_ads4} has zero Ricci scalar $R=0$.% 
\footnote{To see this note that AdS$_4$ has Ricci scalar $-12$ (for unit radius) while $\C \mbox{P}^3$ has Ricci scalar $48$ (for unit radius) hence $-12  + \tfrac{1}{4} 48 = 0$.}
Type IIA string theory on AdS$_4\times \C \mbox{P}^3$ is holographically dual to ABJM theory, a $\CN=6$ superconformal Chern-Simons theory in three dimensions. 
This background has 24 supercharges, and is also exact with respect to $\alpha'$ curvature corrections. Note however that the dictionary relating the physical quantities of ABJM theory and string theory receives corrections \cite{Bergman:2009zh}.

For the above background \eqref{metric_ads4} we have again rescaled the fields giving a dimensionless string tension, corresponding to the rescaled string length $l_s$ given by
\begin{equation}
\label{hat_lambda}
l_s^2 =  \frac{1}{\pi \sqrt{2\hat{\lambda}}}
\end{equation}
Here $\hat{\lambda}$ is related to the 't Hooft coupling $\lambda$ of ABJM theory as $\hat{\lambda}= \lambda - 1/24$. The reason for this shift is that the relation between the 't Hooft coupling and the original radius of the background (above included in the rescaled string length) receives corrections \cite{Bergman:2009zh}. 

In \cite{Ekhammar:2023cuj} Ekhammar et al have generalized the Hagedorn temperature of \cite{Harmark:2017yrv,Harmark:2018red,Harmark:2021qma} to ABJM theory, employing the QSC equations of ABJM \cite{Cavaglia:2014exa}. They found the perturbative expression of $T_{\rm H}$ up to eight loops. Importantly for this paper, they found as well a numerical curve up to $\hat{\lambda}^{1/4} = 2.74$ ($\hat{\lambda}= 56.4$) from which they extracted the asymptotic Hagedorn temperature%
\footnote{The errors in the coefficients are as follows. Third coefficient $\pm 0.0004$, fourth coefficient $\pm 0.003$, fifth coefficient $\pm 0.006$, sixth coefficient $\pm 0.004$ and seventh coefficient $\pm 0.0006$ \cite{Ekhammar:2023cuj}.}
\begin{equation}
\label{TH_AdS4}
T_{\rm H} = \frac{\hat{\lambda}^{1/4}}{2^{5/4} \sqrt{\pi}} + \frac{3}{8\pi} 
- \frac{0.0308}{\hat{\lambda}^{1/4}}+ \frac{0.046}{\hat{\lambda}^{1/2}}- \frac{0.017}{\hat{\lambda}^{3/4}}+ \frac{0.005}{\hat{\lambda}}+ \frac{0.0004}{\hat{\lambda}^{5/4}} + \CO(\hat{\lambda}^{-3/2})
\end{equation}
where the first two coefficients were fixed to their expected exact values in order to get a more accurate fit for the remaining ones.

Finally one can consider  AdS$_3$ with a 7-dimensional product space, meaning 
AdS$_3 \times S^3 \times T^4$, AdS$_3 \times S^3 \times K3$ or AdS$_3 \times S^3 \times S^3 \times S^1$. Thus, we have metric
\begin{equation}
ds^2 =   - (1+r^2) dt^2 + \frac{dr^2}{1+r^2} +r^2 d\phi^{\, 2} +  ds_{7}^{\, 2} \,,
\end{equation}
supported by a three-form Ramond-Ramond flux in part along AdS$_3$ and in part along the seven-dimensional product space $ds_7^{\, 2}$ as well. We use again the parametrization \eqref{string_length}. The product space is such that the Ricci scalar $R=0$. 
The dual field theories are symmetric orbifold CFTs \cite{Strominger:1996sh,Eberhardt:2017pty}.
For these CFTs there are no predictions for the Hagedorn temperature as computed using their integrability symmetries. However, see \cite{Cavaglia:2022xld} for recent progress on obtaining the QSC description of the spectrum in the case of AdS$_3 \times S^3 \times T^4$.

%%%%%%%%%%%%%%%%%%%%%%%%%%%%%%%%%%%%%%%%%
\subsection{Hagedorn temperature from thermal scalar}
\label{sec:ads_thermal_scalar}

We now consider the computation of the Hagedorn temperature from the thermal scalar formalism of Sec.~\ref{sec:thermal_scalar} for the above AdS backgrounds reviewed in Sec.~\ref{sec:ads_backgr}. 

The AdS backgrounds of Sec.~\ref{sec:ads_backgr} can all be written as product geometries AdS$_{d+1} \times M_{9-d}$ with metric
\begin{equation}
\label{met_ads_prod}
ds^2 =  - (1+r^2) dt^2 + \frac{dr^2}{1+r^2} +r^2 d\Omega_{d-1}^2 + dM_{9-d}^{\, 2} \,,
\end{equation}
with constant dilaton, no Kalb-Ramond flux and zero Ricci scalar $R=0$, where $M_{9-d}$ denotes a $(9-d)$-dimensional geometry. Moreover, they are supported by a Ramond-Ramond flux for the AdS space, as well as additional Ramond-Ramond flux for $M_{9-d}$, but there is no flux that mixes the two spaces. 
We choose the flow-of-time vector field 
\begin{equation}
V = \frac{\partial}{\partial t} \,,
\end{equation}
and hence 
\begin{equation}
\label{tau_ads}
\tau = - \frac{i \beta}{2\pi} \frac{\partial}{\partial t} \,.
\end{equation}
Since this is within the AdS part of the product geometry \eqref{met_ads_prod}, the $M_{9-d}$ part of the geometry decouples completely in the thermal scalar equation.

We first recap the results obtained from the leading order thermal EOM given by Eqs.~\eqref{chi_eq}, \eqref{nablasq_L} and \eqref{m_square_L}. Using \eqref{met_ads_prod} and \eqref{tau_ads} we find the EOM
\begin{equation}
\label{leading_ads_EOM}
\frac{1}{r^{d-1}} \partial_r \Big( r^{d-1} (1+r^2) \partial_r \chi \Big) = m^2 \chi \spa m^2 = 
- \frac{2}{l_s^2} + \frac{\beta^2}{4\pi^2 l_s^4} (1+r^2)  \,.
\end{equation}
Here we assume that $\chi$ only depends on the AdS radius $r$ as this is the only coordinate in $m^2$ and we do not expect $\chi$ to break the $SO(d)$ symmetry of the $S^{d-1}$ sphere (and neither the symmetries of the $M_{9-d}$ geometry). 

The leading order Hagedorn temperature $T_{\rm H}$ for $l_s \rightarrow 0$ is simply obtained from $m^2=0$ for $r=0$ giving $T_{\rm H} = \beta^{-1} = \tfrac{1}{2\sqrt{2}\pi l_s}$. Note that $r=0$ since the mode is localized around $r=0$. This reproduces the flat space Hagedorn temperature of type II string theory, in accordance with the reasoning of \cite{Harmark:2018red} in case of AdS$_5 \times S^5$. 

The first order correction to the flat-space Hagedorn temperature was computed in 
\cite{MaldacenaPrivateCommunication,Urbach:2022xzw} using the thermal scalar equation. In this case we consider the leading order approximation of the LHS of the EOM
\begin{equation}
\frac{1}{r^{d-1}} \partial_r \Big( r^{d-1}  \partial_r \chi \Big) = m^2 \chi \,.
\end{equation}
The LHS scales like $1/r^2$ for small $r$ which should in particular match the scaling of the $r^2$ term in $m^2$. This requires that $1/r^2 \sim r^2 / l_s^2$ for small $r$ and $l_s$, hence the mode has $r \sim \sqrt{l_s}$ for $l_s\rightarrow 0$.
As the LHS resembles a $d$-dimensional flat-space Laplacian it seems sensible to assume the mode at lowest orders has a harmonic oscillator ground state profile
\begin{equation}
\chi (r) = \CN \exp \left( - b \frac{r^2}{l_s} \right) \,,
\end{equation}
such that the size of the mode is $r \sim \sqrt{l_s}$ and $b$ is a number that should be fixed from the EOM. We compute from this ansatz
\begin{equation}
\frac{1}{r^{d-1}} \partial_r \Big( r^{d-1}  \partial_r \chi \Big) = \left( - \frac{2  d}{l_s}b +  \frac{4 }{l_s^2}b^2 r^2 \right)  \chi \,.
\end{equation}
It is straightforward to see that the RHS matches $m^2 \chi$ from Eq.~\eqref{leading_ads_EOM} provided
\begin{equation}
b = \frac{1}{\sqrt{2}} + \CO(l_s) \spa \beta = 2 \sqrt{2} \pi l_s - \pi d \, l_s^2 + \CO( l_s^3) \,.
\end{equation}
From this we get the Hagedorn temperature $T_{\rm H}=1/\beta$ including the first order correction as
\begin{equation}
T_{\rm H} = \frac{1}{2\sqrt{2}\pi l_s} + \frac{d}{8\pi} + \CO( l_s ) \,.
\end{equation}
This result was first derived in \cite{MaldacenaPrivateCommunication,Urbach:2022xzw}. Using \eqref{string_length} and $d=4$ this gives $T_{\rm H} = \sqrt{\tfrac{g}{2\pi}} + \tfrac{1}{2\pi} + \CO( g^{-1/2} )$ which fits with the numerical results Eqs.~\eqref{TH_AdS5_HW} and \eqref{TH_AdS5_M} found from the QSC for the Hagedorn temperature of $\CN=4$ SYM \cite{Harmark:2018red,Harmark:2021qma,Ekhammar:2023glu} to the obtained precision, as reviewed in Sec.~\ref{sec:ads_backgr}. Also for $d=3$ it fits with the numerical results of the QSC for the Hagedorn temperature of ABJM theory \cite{Ekhammar:2023cuj}.

After this brief review of known results, we focus now on the higher-order corrections to the Hagedorn temperature, for which our new conjectured 
thermal scalar EOM with curvature corrections, Eqs.~\eqref{m_square_curv} and \eqref{chi_eq_curv}, are relevant. 

To begin, consider the mass squared Eq.~\eqref{m_square_curv}. The $g_{\mu\nu}\tau^\mu\tau^\nu$ is readily found from \eqref{met_ads_prod} and \eqref{tau_ads}. Instead for the $R_{\mu\nu}\tau^\mu\tau^\nu$ term we record the Ricci tensor for the AdS geometry
\begin{equation}
\label{ricci}
R_{\mu\nu} = - d \, g_{\mu\nu} \,.
\end{equation}
Therefore, we obtain
\begin{equation}
\label{ads_m2}
m^2 = - \frac{2}{l_s^2} + \frac{\beta^2}{4\pi^2 l_s^4} (1+r^2) [ 1 - d\, C \, l_s^2 ] \,.
\end{equation}
Again $m^2$ depends only on $r$, thus we assume $\chi$ is a function only of $r$ as well. Using \eqref{ricci} this means we can write Eq.~\eqref{chi_eq_curv} as
\begin{equation}
\label{ads_lapl}
[ 1 + d\, \tilde{C}\, l_s^2 ] \frac{1}{r^{d-1}} \partial_r \Big( r^{d-1} (1+r^2) \partial_r \chi \Big) = m^2 \chi
\,.
\end{equation}
We solve now the EOM for $\chi$ given by Eqs.~\eqref{ads_m2} and \eqref{ads_lapl} perturbatively in an expansion in $l_s$. To this end, we introduce the rescaled radial coordinate
\begin{equation}
\tilde{r} = \frac{r}{\sqrt{l_s}} \,.
\end{equation}
In terms of this rescaled coordinate it is clear from the above discussion that the mode has a finite size for $l_s\rightarrow 0$. 
In terms of this, we write the EOM for $\chi$ in the form
\begin{equation}
\label{general_chi_ads_eq}
\frac{1}{\tilde{r}^{d-1}} \partial_{\tilde{r}} ( \tilde{r}^{d-1} (1+l_s \tilde{r}^2 ) \partial_{\tilde{r}} \chi ) = \Big[ A + 4B^2  \tilde{r}^2 \Big] \chi \,,
\end{equation}
with the identification
\begin{equation}
 \frac{l_s \, m^2}{1 +   d \, \tilde{C} \, l_s^2  } = A + 4B^2  \tilde{r}^2  \,.
\end{equation}
This means
\begin{equation}
\label{ABrel}
 A  = \frac{4B^2-2}{l_s} + 2 d\, \tilde{C} \, l_s + \CO(l_s^3)
\spa 
B = \frac{\beta}{4\pi l_s} \left[ 1 -  \frac{d}{2} (C +\tilde{C}) l_s^2 \right] + \CO( l_s^4) \,.
\end{equation}
Based on this we write the following perturbative ansatz for $A$ and $B$
\begin{equation}
A = a_0 + a_1 l_s + a_2 l_s^2  + \CO(l_s^3) \spa
B = \frac{1}{\sqrt{2}} + b_1 l_s + b_2 l_s^2 + b_3 l_s^3 + \CO(l_s^4) \,,
\end{equation}
where the first coefficient of $B$ corresponds to the zeroth order solution found above, and we expanded $A$ and $B$ to the order determined by the thermal scalar EOM with curvature corrections, Eqs.~\eqref{m_square_curv} and \eqref{chi_eq_curv}.

Solving \eqref{general_chi_ads_eq} in general for $A$ given $B$, and requiring $\chi(\tilde{r})$ to be a normalizable mode, fixes the $a_n$ coefficients uniquely in terms of the $b_n$ coefficients as follows
\begin{equation}
\label{gen_a_coeff}
a_0=-\sqrt{2}d\spa a_1 = - \frac{d^2+2d +8b_1 d}{4} \spa a_2 = \frac{\sqrt{2} d^2 + 2\sqrt{2} d- 64 b_2 d}{32}  \,.
\end{equation}
We note that $\chi(\tilde{r})$ has the normalizable form
\begin{equation}
\chi(\tilde{r}) =  \CN \left[ 1 + \sum_{n=1}^\infty l_s^{\, n} \sum_{m=1}^{2n} c_{n,m} \tilde{r}^{2m}  \right] \exp \left( - \frac{\tilde{r}^2}{\sqrt{2}} \right) \,,
\end{equation}
where one should include up to $l_s^2$ corrections in the above procedure, which fixes $c_{1,m}$ and $c_{2,m}$ in terms of the $b_n$ coefficients.

Combining now \eqref{ABrel} with \eqref{gen_a_coeff} we find the $b_n$ coefficients
\begin{equation}
b_1 = -\frac{d}{4} \spa b_2 = - \frac{d(1+4\tilde{C})}{8\sqrt{2}} \spa b_3 = \frac{d(2+d)}{128}
\end{equation}
With this, we find $T_{\rm H}=1/\beta$ from \eqref{ABrel} as
\begin{equation}
\label{TH_gen_d}
T_{\rm H} = \frac{1}{2\sqrt{2}\pi l_s} + \frac{d}{8\pi} + \frac{d^2-4d \, C +4}{16\sqrt{2} \pi} l_s 
+ \frac{4d^3 + 7d^2 - 2d + 16 (\tilde{C}-C) d^2 }{256 \pi} l_s^2 + \CO( l_s^3 )
\end{equation}
as the prediction of the Hagedorn temperature for type II string theory on AdS$_{d+1}$ backgrounds.

For $d=4$ and in terms of $g$ defined by \eqref{string_length}, this becomes
\begin{equation}
\label{TH_d4}
T_{\rm H} = \sqrt{\frac{g}{2\pi}} + \frac{1}{2\pi} + \frac{5-4C}{8 \pi \sqrt{2\pi g} } + \frac{45 + 32(\tilde{C}-C)}{128\pi^2 g} 
\end{equation}
In Sec.~\ref{sec:ppwave_thermal_scalar} we fixed $C=2\log 2$ by matching the curvature correction of thermal scalar EOM to certain pp-wave backgrounds in which the Hagedorn temperature is known exactly. With this value of $C$, we obtain the numerical value of the $1/\sqrt{g}$ coefficient of $T_{\rm H}$ to be $- 0.00865382$ which fits with the numerical results \eqref{TH_AdS5_HW} and \eqref{TH_AdS5_M} obtained from the QSC for $\CN=4$ SYM in \cite{Harmark:2018red,Harmark:2021qma} and \cite{Ekhammar:2023glu}, respectively.
This is a non-trivial result in that it shows the coefficient of the curvature term in $m^2$ given by Eq.~\eqref{m_square_curv} is consistent with both the pp-wave and AdS backgrounds.%
\footnote{Note that while the flow-of-time vector field for the AdS$_{d+1}$ backgrounds only probes the AdS geometry, the  flow-of-time vector field for the pp-wave backgrounds of Section \ref{sec:ppwave} arise from a limit that includes the product space. {\sl E.g.}~for AdS$_5 \times S^5$ the flow-of-time vector field for the corresponding pp-wave background includes the $S^5$.}

We notice now, that unlike the pp-wave backgrounds of Section \ref{sec:ppwave}, the AdS backgrounds are sensitive to the $\tilde{C}$ coefficient. Choosing $C = \tilde{C} = 2\log 2$, as written in Eq.~\eqref{theCs},
%
%\begin{equation}
%\label{theCs}
%C = \tilde{C} = 2\log 2
%\end{equation}
%
means that the $l_s^2$ coefficient in \eqref{TH_gen_d} does not receive corrections from the curvature terms. For the $d=4$ case given by \eqref{TH_d4} one finds the numerical value of the $1/g$ coefficient of $T_{\rm H}$ to be $0.035607$ which again is consistent with both \eqref{TH_AdS5_HW} and \eqref{TH_AdS5_M}. 
However, we note here that $\tilde{C}$ is not fixed to be an exact value by this comparison, unlike $C$ which is fixed exactly by comparing to the pp-wave backgrounds in Section \ref{sec:ppwave}.

Turning to $d=3$ we find using \eqref{hat_lambda} 
\begin{equation}
T_{\rm H} = \frac{(2\hat{\lambda})^{1/4}}{2 \sqrt{2\pi}} + \frac{3}{8\pi} 
- \frac{6\log 2 - 3}{4 \sqrt{2} \pi^{3/2} (2\hat{\lambda})^{1/4}} 
+ \frac{165}{256\pi^2 \sqrt{2\hat{\lambda}}} + \CO( \hat{\lambda}^{-3/4} )
\end{equation}
where we inserted the values given by \eqref{theCs} for $C$ and $\tilde{C}$. With this, one obtains $-0.3093$ for the $\hat{\lambda}^{-1/4}$ coefficient and $0.0462$ for the $\hat{\lambda}^{-1/2}$ coefficient which fit with the QSC result \eqref{TH_AdS4} of \cite{Ekhammar:2023cuj}.

For general $d$, the Hagedorn temperature \eqref{TH_gen_d} is same as the conjectured analytic expression of \cite{Ekhammar:2023glu,Bigazzi:2023hxt,Ekhammar:2023cuj} in accordance with the analysis of the zero-mode contribution to the string spectrum in \cite{Bigazzi:2023hxt}. However, in contrast with the above,  the result for $T_{\rm H}$ of \cite{Ekhammar:2023glu,Bigazzi:2023hxt,Ekhammar:2023cuj} is based on the following conjectured ansatz for the thermal scalar EOM \cite{Ekhammar:2023glu}
\begin{equation}
\label{ads_M}
 \frac{1}{r^{d-1}} \partial_r \Big( r^{d-1} (1+r^2) \partial_r \chi \Big) = M^2 \chi
 \spa M^2 =  \frac{\beta^2}{4\pi^2 l_s^4} (1+r^2)- \frac{2}{l_s^2} + \frac{\beta^2}{2\pi^2 l_s^2} \Delta c
 \end{equation} 
which can reproduce the same result \eqref{TH_gen_d} provided $\Delta c = -d \log 2$. From the discussion of Sec.~\ref{sec:thermal_scalar_curv} it is clear that the ansatz \eqref{ads_M} cannot arise from a general curvature correction term to the mass square of $\chi$. In particular, the presence of $\beta^2$ necessarily has  to occur together with $-g_{tt}= 1+r^2$ to take into account the proper length of the thermal circle. This makes it necessary to include an $r$ dependent curvature correction to the mass square of $\chi$ in contrast with $M^2$ in \eqref{ads_M}. Therefore, we conclude that the ansatz \eqref{ads_M} cannot be correct.

%%%%%%%%%%%%%%%%%%%%%%%%%%%%%%%%%%%%%%%%%
\section{Discussion}
\label{sec:discussion}

In  this paper we have proposed (see Section \ref{sec:thermal_scalar}) a curvature-corrected thermal scalar EOM that should be able to predict the Hagedorn temperature of type II string theory, at least for backgrounds that do not receive $\alpha'$ corrections. 
Without Kalb-Ramond flux, the curvature-corrected thermal scalar EOM is given by 
Eqs.~\eqref{m_square_curv} and \eqref{chi_eq_curv}. Also turning on Kalb-Ramond flux, it is given by 
Eqs.~\eqref{chi_eq_curv_B} and \eqref{m_square_curv_B}.
In Section \ref{sec:ppwave} we have checked that our proposed curvature-corrected thermal scalar EOM, with coefficients \eqref{theCs}, provides the correct predictions for a family of pp-wave backgrounds with Ramond-Ramond flux and a pp-wave with both Ramond-Ramond and Kalb-Ramond flux for which one has an exact calculation of the Hagedorn temperature from the string spectrum. Moreover, in Section \ref{sec:ads} we have checked  the proposed thermal scalar EOM for two holographic AdS backgrounds for which one has numerical predictions from the QSC technique.

Comparing the expansion \eqref{ppwave_TH_eq2} of the exact Hagedorn temperature for pp-wave backgrounds considered in this work to the expansion \eqref{TH_ppwave_curv}, found from the curvature-corrected thermal scalar EOM, we notice that there are additional higher-order terms in $\beta f$, starting from the following term of order $\beta^4f^4$
\begin{equation}
\label{fourth_order}
- \frac{3\zeta(3)}{32\pi^3} (a_1^{\, 4}+a_2^{\, 4}) \beta^4 f^4
\end{equation}
Can one reproduce this term from the thermal scalar EOM? This is an interesting question as there are only few possible non-zero curvature correction terms for the pp-wave background that can contribute to this order. This is due to the fact that $g^{++}=0$, that the only non-zero component of the Ricci tensor is $R_{++}$ and the non-zero components of the Riemann curvature tensor are listed by \eqref{Riemann_ppwave}. At the same time, $\beta^4 f^4$ can only arise from a contraction of four $\tau^\mu$ with a curvature tensor which is quadratic in the Riemann curvature tensor with four free indices. The only two possible non-zero correction terms that one can add to $m^2$ are therefore
\begin{equation}
\label{curv_corr_K}
 \frac{K}{l_s^2} R_{\mu \alpha \nu \beta } R_{\rho}{}^\alpha{}_\sigma{}^\beta \tau^\mu \tau^\nu \tau^\rho \tau^\sigma + \frac{\tilde{K}}{l_s^2} (R_{\mu\nu} \tau^\mu \tau^\nu)^2
\end{equation}
However, the second one of these gives a contribution to the expansion \eqref{TH_ppwave_curv} which goes like $(a_1^2 +a_2^2)^2 \beta^4 f^4$. Thus we need $\tilde{K}=0$. Instead the first one is proportional to $(a_1^4+a_2^4)\beta^4f^4$. If we choose the coefficient 
\begin{equation}
K = \frac{3}{2}\zeta(3)
\end{equation}
we reproduce correctly the term \eqref{fourth_order} in the expansion  \eqref{ppwave_TH_eq2}. It would be interesting to explore this further.

A related future direction is to study the higher-order corrections for AdS spaces as well, starting from order $l_s^4/\mathcal{R}^4$. For the AdS backgrounds there are additional possible curvature correction terms in addition to \eqref{curv_corr_K} that one can add to the thermal scalar EOM and which are zero for the pp-wave backgrounds. To this end, it would be useful obtaining higher precision especially for the expansion coefficients \eqref{TH_AdS4} for the AdS$_4\times \C \mbox{P}^3$ pioneered in \cite{Ekhammar:2023cuj}, as well numerical predictions from one of the AdS$_3$ backgrounds mention in Section \ref{sec:ads_backgr}.

Relatedly, one could explore adding chemical potentials for the AdS spaces. 
This could potentially be important for exploring for higher-order corrections at order $l_s^4/\mathcal{R}^4$ or higher, as one could possibly use the chemical potentials to distinguish the possible curvature terms. 
In \cite{Harmark:2021qma} it is described how to extend the QSC technique to include chemical potentials for the dual integrable superconformal field theory. However, so far no predictions of the Hagedorn temperature at strong coupling has been put forward.
On the string side, it is described in Section \ref{sec:ppwave_chempot} how to include a chemical potential for the thermal scalar in the case of pp-wave backgrounds. It would be interesting to use the thermal scalar more generally to predict the dependence of the Hagedorn temperature on a chemical potential. 

We have seen that the curvature-corrected thermal scalar EOM of this work successfully can predict the Hagedorn temperature for several important type IIB string theory backgrounds supported by Ramond-Ramond flux. Thus, it seems to be a valid effective theory for the thermal scalar mode. Therefore, as a final comment, 
it would be interesting to derive the corrected thermal scalar EOM  of this work by a first-principle derivation in a low energy expansion of string theory. Here are two important points:
\begin{itemize}
\item Previous works have considered tachyon fields in the NS-NS sector (see for example \cite{Tseytlin:1993df}). However, the curvature corrections to $\chi$ at first order in curvature require the presence of Ramond-Ramond flux, in which case string theory is generally more difficult to quantize. 
\item A difference from previous works involving tachyon fields is that the tachyonic nature of $\chi$ arise due to supersymmetry breaking boundary conditions for a particular isometry in the Euclidean section, {\sl i.e.}~the thermal circle. This is why $m^2$ depends on $\tau^\mu$ in the Lorentzian formulation. 
\end{itemize}

%%%%%%%%%%%%%%%%%%%%%%%%%%%%%%%%%%%%%%%%%
\section*{Acknowledgments}

TH thanks Niels Obers, Bo Sundborg, Arkady Tseytlin and Kostya Zarembo for interesting and stimulating discussions and comments. TH also thanks an anonymous referee of JHEP for very useful feedback.

%%%%%%%%%%%%%%%%%%%%%%%%%%%%%%%%%%%%%%%%%%%%%%%%%%%
%%%%%%%%%%%%%%%%%%%%%%%%%%%%%%%%%%%%%%%%%%%%%%%%%%%
%%%%%%%%%%%%%%%%%%%%%%%%%%%%%%%%%%%%%%%%%%%%%%%%%%%
\addcontentsline{toc}{section}{References}

%\bibliography{mybib}
%\bibliographystyle{newutphys}

\providecommand{\href}[2]{#2}\begingroup\raggedright\endgroup

\end{document}